# Chikungunya: The Silent Threat in the Shadows

Ambreen Talib[1], Rabbya Rayan Shah[1], Rameen Atique[1], Hafiza Arshi Saeed[1], Ayesha Haidar[1], Ayesha Nadeem[1], Areesha Naveed[1], Javeria Sharif[1], Abdul Samad[2]

[1]Department of Pathobiology and Biomedical Sciences, FV&AS, Muhammad Nawaz Shareef University of Agriculture, Multan, 25000, Pakistan

[2]Division of Applied Life Science (BK21 Four), Gyeongsang National University, Jinju 52852, Korea

**Corresponding Author:**

Abdul Samad

**Email:** buzdarabdulsamad@gmail.com

**Abstract**

Chikungunya virus (CHIKV) is currently one of the most relevant arboviruses to public health. It is a member of the *Togaviridae* family and *alphavirus* genus and causes an arthritogenic disease known as Chikungunya fever (CHIKF). It is a multifaceted disease that is distinguished from other arbovirus infections by the intense arthralgia that can last for months, even years, in some individuals. The virus has re-emerged as a global health threat in recent decades, spreading from its origin in Africa across Asia and America, leading to widespread outbreaks impacting millions of people. Despite more than 50 years of research on CHIKV pathogenesis, there are still no drugs or vaccines available. Current management involves providing supportive care to alleviate symptoms and improve the quality of patients' lives. The ongoing threat of CHIKV shows that there is an intense need to understand its pathogenesis better. It is crucial to comprehend the mechanism underlying the disease for the development of prevention and control measures. This review aims to provide a comprehensive overview of CHIKV, shedding light on host factors, vector-related factors, and complex interactions of viral genetics. By seeking these intricate connections, the review aims to communicate the knowledge about CHIKV, offering insights that may ultimately lead to more effective strategies for the prevention and management of this re-emerging global health threat.

## 1. Introduction:

Chikungunya virus (CHIKV) is a mosquito-transmitted virus that belongs to the Semliki Forest antigenic group of the Togaviridae family and Alphavirus genus, which has been responsible for recurring epidemics over the years globally [1]. Its genome is closely related to that of the o'nyong-nyong virus. CHIKV is a positive sense, single-stranded RNA (ssRNA) virus with an 11.8kbp size of the genome, which consists of 1244 amino acids structural and 2472 amino acids nonstructural polyprotein [2], which is responsible for causing arthritogenic disease worldwide, resulting in a significant public health threat [3]. The virus is primarily transmitted to humans via the bite of an infected mosquito from the genus *Aedes* spp., mainly *Aedes aegypti* and *Aedes albopictus*, which are highly domesticated and adaptable to extreme environmental changes, therefore resulting in an efficient spread in tropical and subtropical regions across the countries and continents [4, 5].

Chikungunya virus infection results in a disease known as chikungunya fever (CHIKF). It is characterized by high fever, headache, skin rash, incapacitating arthralgia, prominent polyarthralgia, and myalgia that can last week's to months even years in some individuals after complete virus clearance, resulting in notorious social and economic impact [6, 7, 8]. The word chickungunya means, "To walk bent over or to become contorted" in the African dialect Makonde and refers to the remarkable characteristics of this disease like intense and persistent joint pain seen in the affected patients [9]. CHIKF is a non-deadly disease but atypical and severe acute manifestations can cause multiple organ failure and death. Its mortality rate is 0.024 up to 0.7%, which depends upon the virus genotype and the neurological system condition [10, 11]. Its first case was reported in Tanzania in 1953 in a febrile individual [12]. In 1958, it was reported in Bangkok and Thailand [13]. The first known emergence of CHIKV occurred in the Southwestern Indian Ocean region in April 2005. By the end of 2005, apparently with a gap of 32 years during which CHIKV was not reported, India reported this CHIKF in numerous states, with suspected cases number reaching more than 1.3 million. CHIKV continued to spread and cause outbreaks in Sri Lanka and other countries in Southeast Asia. In December 2013, the World Health Organization (WHO) reported the first local transmission of CHIKV in the Western Hemisphere on the Caribbean island of Saint Martin [14]. In less than 10 years, CHIKV has spread from the coast of Kenya to Caribbean regions and has caused millions of cases in over 50 countries. In other words, CHIKV has reemerged as a true global pathogen [15].

Despite the reemergence and relevance of CHIKV to public health, there is still no vaccine or antiviral drug available for the prevention and treatment of CHIKF [16]. In contrast to Zika virus (ZIKV) and Dengue virus (DENV), in which experimental models are known and used, there is no model for studying CHIKV infection as it is diverse and not rarely reproduces just a piece of pathogenesis observed in humans, which is challenging for development of vaccines and drugs [17]. Therefore, understanding this complex disease, as well as the current laboratory limitations, is vital to address CHIKF [18]. In this review, the major points of CHIKV epidemiology, replication, and pathogenesis are discussed. It also discusses the priorities for further studies needed for effective disease control, management, and prevention of this reemerging global health threat.

## 2. Epidemiology:

In 1823, the first case of a chikungunya-like disease was reported in humans in Zanzibar, Africa, followed by the epidemic of a similar disease on Thomas Island in the Caribbean between 1827 and 1828 [19]. Since then, no other cases of related diseases were identified until 1952, when rheumatic fever affected many people in Tanzania. This was the first time CHIKV was identified, isolated, and characterized as an *arbovirus*. Then, CHIKV caused local and sporadic outbreaks in Asia and Africa until 2004, and it has spread to over 60 countries all over the world [20, 21]. Between 2005 and 2006, most remark outbreaks occurred on La Reunion Island in the Indian Ocean. During this epidemic, 260000 people were affected, with an average of 40000 new cases every week and 284 deaths [22]. This epidemic was caused by a new vector, A. albopictus, that helped in virus propagation [23]. This vector is highly adaptable to temperate regions. CHIKF was reported in America in 2013 [24] when the outbreak was observed in Saint Martin, with 658 confirmed cases and a 1.76% infection rate [25]. Then, 45 countries were affected, and more than 3 million cases were observed [26, 27]. It spread in Florida in July 2014. Mosquito larvae and egg transport by ships and air traffic have also been described as means of disseminating mosquitoes into naive and suitable environments [28]. Two interconnected transmission pathways take place; these include urban and sylvatic cycles. In the sylvatic cycle, transmission occurs between forest-dwelling Aedes mosquitoes and non-human primates, resulting in small outbreaks and sporadic human cases [29, 30]. The urban cycle is related to transmission via *Aedes aegyptae* and *Aedes albopictus* from infected to non-infected individuals [31, 32]. Enhanced infectivity and transmission of CHIKV by A. albopictus is due to Ala-Vla mutation at position 226 in the protein E1 gene (E1:A226V) of an ECSA lineage strain abolished virus dependence on cholesterol to replicate, which is also important in virus spread to different continents [33, 34, 35]. In addition to the urban and sylvatic cycles of transmission, CHIKV is also transmitted by vertical transmission during pregnancy and blood transfusion. This route is a signal of alert for uncontrolled transmission and the potential risk of pandemics [36]. Vertical transmission was observed during pregnancy, and CHIKV infection in neonates is diverse and varies from asymptomatic to severe, in which meningoencephalitis and myocarditis are the most common signs of severity [37]. Suppose the mother is under a viremia period during childbirth. In that case, there is an increased risk for the development of severe symptoms in neonates, but this risk softens if the infection occurs 4 weeks before the birth [38, 39]. It has been observed that CHIKV infection can be spread by blood transfusion or transplantation during its outbreak. Extra care should be taken during blood transfusion in places where CHIKV is endemic or outbreaks are ongoing because its prevalence ranges from 0.4-2.1% in blood during epidemics [40, 41].

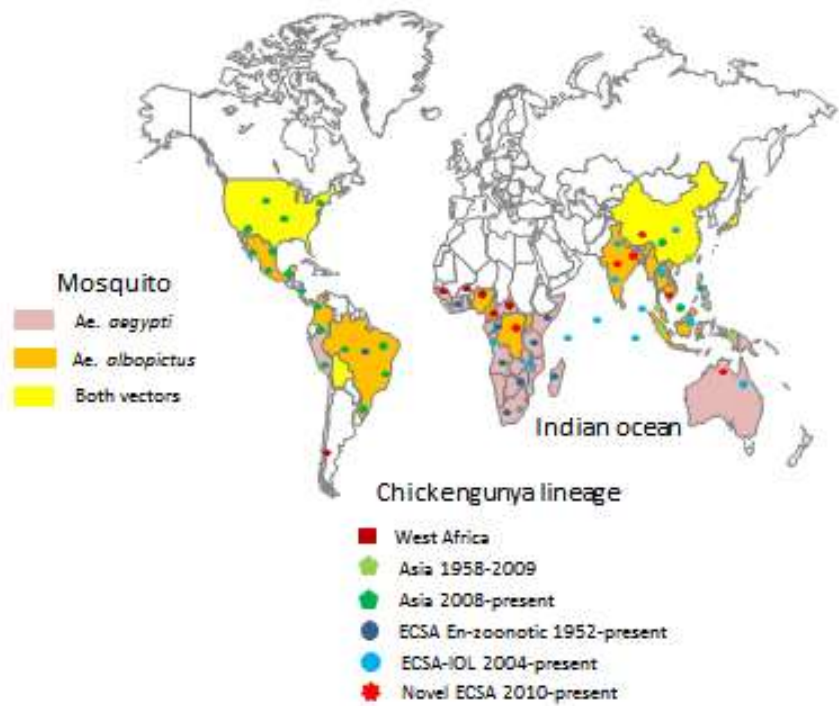

**Figure 1: Geographical distribution of CHIKV lineages during 1952–2024 [42, 43, 44, 45, 46]**

## 3. Pathogenesis:

A great deal of research is being done on the pathogenesis of CHIKV infections, and small animal and nonhuman primate models of acute and chronic illness have recently been produced [47]. Research on humans and models of animals has demonstrated that CHIKV infection of cells in musculoskeletal tissues, such as fibroblasts and osteoblasts, and infiltration of inflammatory cells, primarily monocytes, macrophages, natural killer cells, and T cells, are linked to disease signs and symptoms after infection. Damage to the musculoskeletal system during the acute phase of CHIKV infection [48, 49]. Indicating that CD4T cells have a harmful function in CHIKV illness. Furthermore, research using a well-established mouse model of infection with the similar arthritogenic alphavirus Ross River virus revealed that substances produced by virus-infected osteoblasts increase the recruitment of monocytes into joints, which in turn promotes the development of arthritis [50].

High levels of interleukin-6 (IL-6), IL-1, RANTES, monocyte chemoattractant protein 1 (MCP-1), monokine induced by gamma interferon (MIG), and IP-10 are linked to CHIKV disease severity in humans. CHIKV sickness is also associated with higher serum levels of particular cytokines and chemokines [51]. Crucially, there is minimal evidence that autoimmunity develops in people with chronic illness, and the reason for persistent CHIKV joint dysfunction is unknown [52]. Given that persistent arthritis has been linked to increased levels of IL-6 and granulocyte-macrophage colony-stimulating factor (GM-CSF), cytokines may also play a role in chronic CHIKV illness [53]. Furthermore, those with persistent symptoms may have modestly raised C-reactive protein (CRP), which indicates persistent inflammation. A chronic CHIKV infection in the musculoskeletal tissues may lead to chronic CHIKV joint dysfunction. RNA and CHIKV antigens were found in synovial tissue biopsy samples taken from a patient with persistent joint discomfort [54]. Additionally, in a muscle tissue biopsy specimen taken from a patient during a relapse of persistent musculoskeletal pain, CHIKV antigen was found in muscle satellite cells [55]. Animal models have also demonstrated the persistence of CHIKV RNA and antigen in tissues [56, 57], which adds more evidence that CHIKV causes long-lasting infections that could encourage immune-mediated chronic illness.

## 4. Transmission:

The chikungunya virus mainly spreads to humans through the bite of infected mosquitoes, primarily *Aedes aegypti* and *A. albopictus* [58]. During epidemic periods, humans serve as the primary hosts for the virus. Mosquitoes become infected when they feed on an individual who is already carrying the virus [59].

### 4.1. Blood-borne transmission:

Bloodborne transmission of the chikungunya virus can occur; there have been documented cases among laboratory personnel handling infected blood and healthcare providers drawing blood from infected patients [60].

### 4.2. In utero transmission

In utero, transmission of the chikungunya virus is rare but has been reported, primarily during the second trimester. Intrapartum transmission has also been observed when the mother was viremic at the time of delivery [61]. The chikungunya virus has not been detected in breast milk, and there are no reported cases of infants contracting the virus through breastfeeding. Given that the benefits of breastfeeding likely outweigh the risks of chikungunya virus infection in nursing infants, mothers should be encouraged to continue breastfeeding, even if they are infected or live in areas with active virus transmission [62, 63].

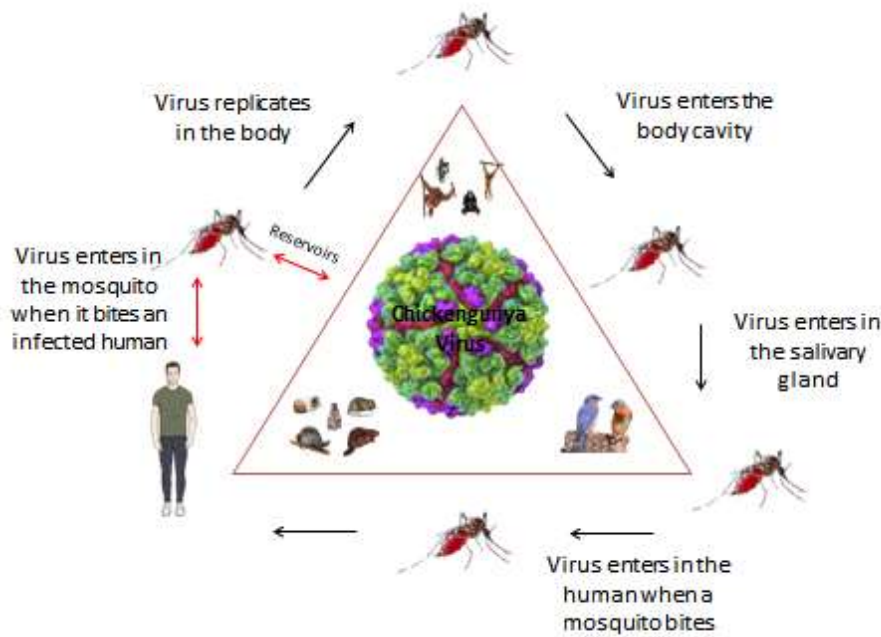

**Figure 2**: **Transmission cycle of Chikungunya Virus**

## 5.  Viral Replication:
### 5.1. Genome structure and organization:

CHIKV contains a single-stranded positive-sense RNA (ssRNA) genome of ~12kb size containing a 5' N7-methylguanylated cap and a 3' polyadenylated tail [64, 65] (fig). Two open reading frames (ORFs) that code for structural proteins (SP) and non-structural proteins (nsP) are present in 5' and 3' untranslated regions of the genome [66, 67]. nsP1, nsP2, nsP3, and nsP4 are present in 5' ORF that play an important role in genome replication and other genome functions and is translated as a polyprotein of nsP1-3 or nsP1-4 by the code of an opal stop codon between nsP3 and nsP4 [68, 69, 70]. Negative sense antigenome transcribes a subgenomic RNA (sgRNA) that is translated into 3' ORF; it encodes for a polyprotein that is composed of the viral sPs (capsid, E1, E2, E3, 6K, and TF), which are the actual constitutes of the virus particles and help in virion assembly, budding, and entry into the host cell [71, 72]. Both polyproteins are processed by proteases of the host or virus during virus replication with spatial or temporal patterns [73, 74]. CHIKV is encapsulated in an icosahedral nucleocapsid shell that is surrounded by a lipid bilayer envelope and is about 70nm in diameter. E1 and E2 glycoproteins' heterodimers form trimeric spikes that are embedded in viral membranes and form an outer icosahedral protein lattice [75].

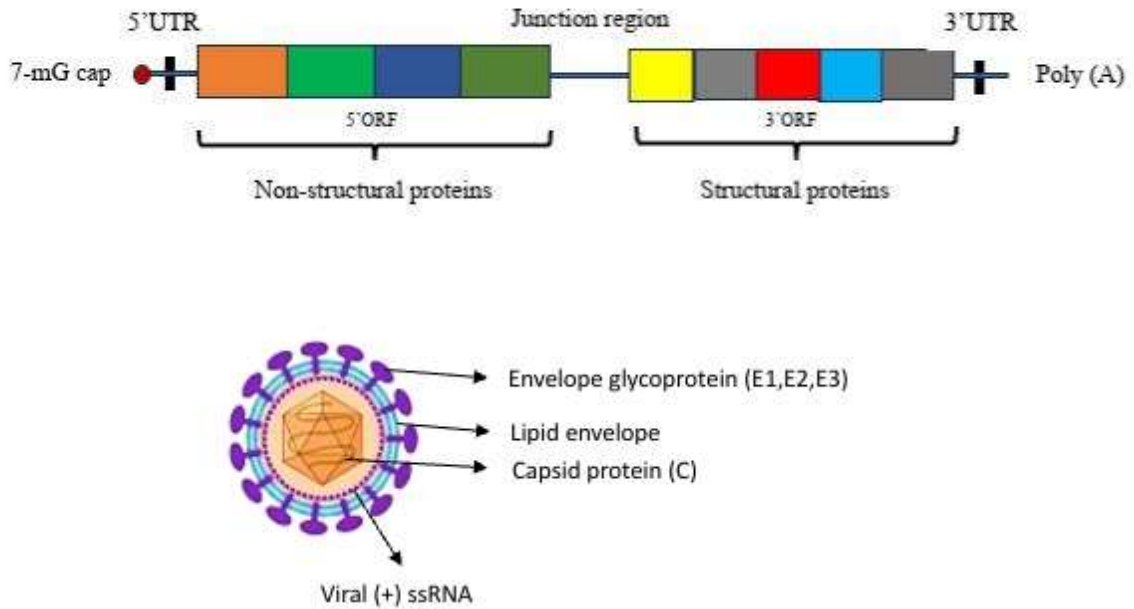

**Figure 3: Genome and structure of Chikungunya virus**

## 5.2. Viral entry:

When CHIKV is inoculated to the host organism, clathrin-mediated endocytosis by plasma membrane occurs due to the binding of E2 glycoprotein to the membrane receptor Mxra 8 on the target cells, which helps in activating the internal signaling pathway that results in the commitment of clathrin molecules to the plasma membrane of the host [76, 77]. However, Mxra8 is not expressed on all target cell types [78, 79], and when Mxra8 KO was infected with CHIKV, virus replication was reduced, and the mice were not abrogated [80], which indicates that CHIKV binds to additional receptors *in vivo*. Then, clathrin molecules are separated from the endocytic blister, and the acidification of endosomal pH triggers the detachment of E1 and E2 heterodimers. In addition, the role of E1 and E2 in the recognition of target cells and fusion of membrane [81]. It also identifies other membrane proteins, including fuzzy homolog protein (FUZ) and tetraspanin membrane protein (TSPAN9), that are required for proper infection process by involving in clathrin-mediated endocytosis and virus entry (i. Virus orientation to the early endosome or ii. Modulation of the endome membrane is more important for the fusion process) respectively [82]. Glycosaminoglycans (GAGs), Dendritic Cell-Specific Intracellular Adhesion molecule-3-

Grabbing Non-integrin (DC-SIGN), membrane protein complex CD147, T-cell immunoglobulin and mucin (TIM) family, and AXL receptor tyrosine kinase also acts as alternative cell receptors for CHIKV [83, 84]. Entry through micropinocytosis may occur through the uptake of apoptotic bodies harboring CHIKV particles from neighboring infected cells [85]. Other infection pathways occur in muscle and epidermal cells in the first cell line. CHIKV follows an eps15-dependent pathway to enter by epidermal growth factor receptor substrate 15. In the second cell line, micropinocytosis is the preferred path, which shows the adaptive evolution of CHIKV to infect the target host cell as another means than clathrin-mediated endocytosis [86, 87]. Only ATP synthase beta-subunit [88] and Heat Shock Chaperone [89] were reported as host factors that help in CHIKV entry into mosquito cells in culture; otherwise, it is difficult to identify entry factors for mosquitoes. CHIKV entry is a complex process due to strains of CHIKV that vary in their dependence on some entry factors [90]; they also show different pH thresholds [91] and cholesterol requirements, thereby, utilize the entry pathways differently [92] and have the ability to infect other cell lines [93, 94]. So, CHIKV entry into different cell types is a complex mechanism that yet has to be completely unraveled.

### 5.3. Genome replication:

In mammalian cells, the CHIKV replication cycle has not been fully understood. Still, recent investigations have added new and exciting details to the viral replicase and other functions of the viral proteins. After membrane fusion, virus nucleocapsid is released into the cytoplasm [95], where interaction between a ribosome binding site in the N-terminus of capsid and the host 60s ribosomal subunits triggers rapid disassembly of the nucleocapsid (NC) and deliver the RNA into the cytosol [96], and protein C detached from the gRNA which is now translated into the nonstructural polyprotein P1, P2, P3, and P4. The nsP polyprotein is directly translated and cleaved into the protein by the protease action of nsP2 [97]. Highly conserved opal stop codon right after nsP3 is present in most of the CHIKV strains that allow for the translation of nsP1-2-3 with low levels of nsP4 [98, 99, 100]. Arginine residues replace opal codon in some strains, and the polyprotein is translated as nsP1-2-3-4, and nsP2's protease action immediately releases nsP4 from P1-3-4 [101]. Viral replicase is directed towards cholesterol-rich microdomains of the plasma membrane by nsP1, where the negative sense strand is synthesized by nsP1-2-3 and host proteins [102]. Guanyltransferase and guanine-7-methyltransferase (mediate RNA capping) initiate negative-strand synthesis by nsP1. nsp2 has NTPase, RNA 5' triphosphatase, and helicase activities that help nsP4 RNA -dependent-RNA-polymerase (RdRP) to synthesize full-length RNA antigenome [102]. Antigenome has a subgenomic promoter that initiates the transcription of the CHIKV sgRNA encoding the sP that helps in viral assembly. The subgenomic RNA is also capped and polyadenylated and is translated into a polyprotein [103].

### 5.4. Virion assembly and release:

Once sP is translated from sgRNA, the autoproteolytic action of the capsid releases it from another sP polyprotein into the cytosol. Capsid interacts with sequences in newly synthesized RNA and promotes oligomerization of capsid molecules and genome encapsidation [104]. The E3-E2-6K or TF-E1 polyproteins are directed to the endoplasmic reticulum (ER) of the host to the signal sequence of the E3 N-terminus [105]. TF is produced by the frameshift of the ribosome on the 6K sequence, followed by the stop codon that stops the translation of E1 [106]. pE2 and E1 undergo post-translational modification (N-li ked glycosylation and palmitoylation) to produce a non-fusogenic and immature spike complex [107]. By succeeding in these events, the nucleocapsid core moves to membrane regions that are rich in E1 and E2 dimers, and a budding mechanism releases mature virion from the infected cell. The budding mechanism is not fully understood, but some studies showed that it is highly dependent on pH and optimal temperature conditions, as well as the presence of the host cell membrane [108]. Moreover, virus release is increased by the presence of 6K and TF proteins as deletion or mutations in their genes decrease the rate and efficiency of virion budding that, indicates their relationship with the process [109]. CHIKV can also be transmitted to neighboring uninfected cells through virally induced Intracellular long extensions (ILEs; > 10uM) (142,143,145,246,149,150), which show active viral budding of infectious viral particles along the length of the extension and concentrated at the top [110].

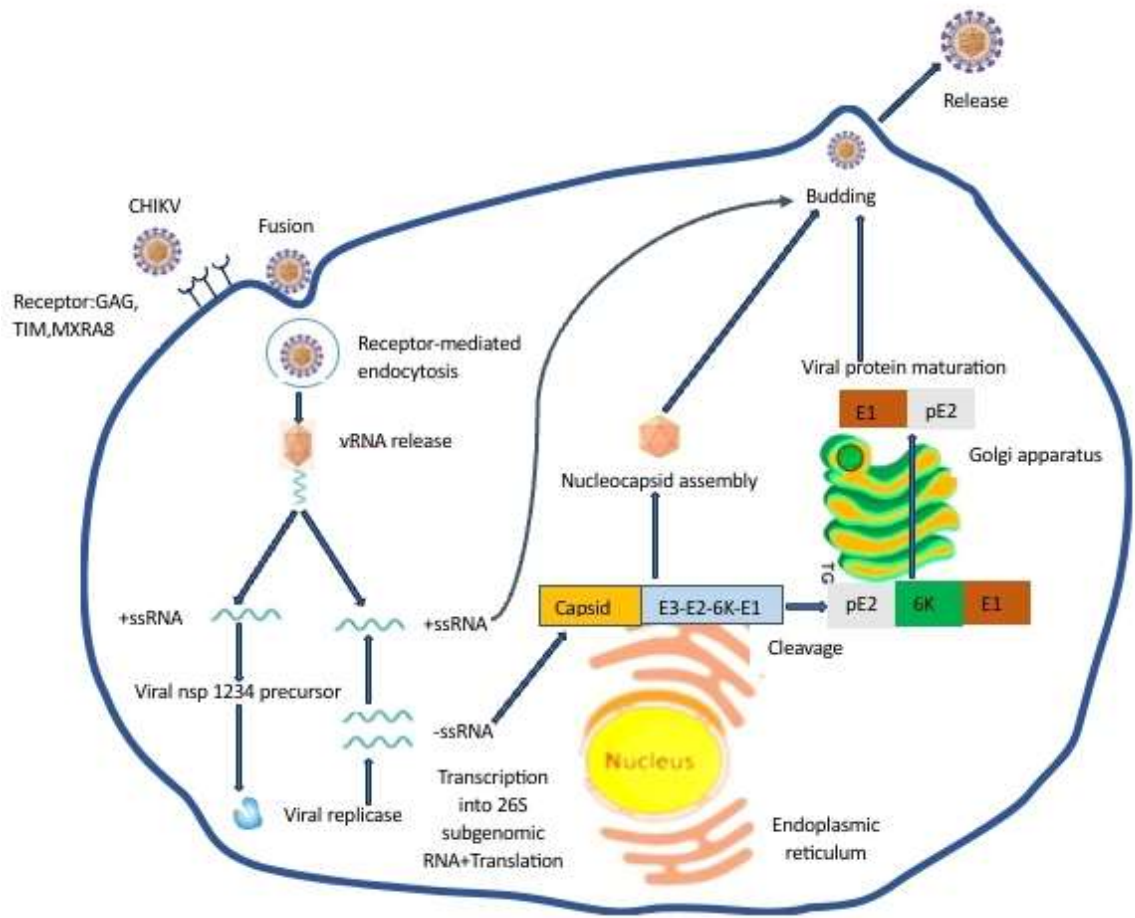

**Figure 4: Chikungunya virus replication cycle**

6. **Clinical manifestations:**

Chikungunya, meaning "disease that bends up the joints," is typified by a sudden onset of fever and excruciating joint pain that can last for weeks or even years [111]. Only 5 to 25% of CHIKV infections are asymptomatic, in contrast to infections with many other arboviruses. Usually symmetrical, arthritis mostly affects peripheral joints such as the wrists, knees, ankles, and tiny hand joints. Other indications of the disease include myalgia, especially in the lower back and leg muscles, tenosynovitis, skin rash, and arthritis, which frequently causes joints to enlarge and become sensitive. Along with these clinical characteristics, CHIKV infection has also been linked to serious neurologic and cardiac symptoms, as well as, in rare cases, fatalities. Patients over 65, those with underlying medical issues, and neonates are more likely to experience these more severe consequences [112]. The substantial public health concern posed by CHIKV is highlighted by the fact that major epidemics have severe economic effects and that chronic CHIKV disease can be quite debilitating [113].

7. **Diagnosis**

Specific anti-CHIKV IgM and IgG on blood samples and the virus's confirmation on initial tests are necessary for diagnostic purposes. CHIKV diagnosis has been accomplished by the development of numerous quantitative reverse transcription-polymerase chain reactions (RT-PCR). There are generic kits that can have exceptional sensitivity and specificity [114]. CHIKV-RNA is found in plasma samples within the first week following the first appearance of symptoms, usually when viremia is at very elevated levels [115]. Other tissues and fluids, such as corneas or other graft tissues, can also be screened using RT-PCR. Additionally, early samples on Vero or C6/36 cell lines can be used to isolate CHIKV. Because this approach may only be employed in Biosafety Level 3 laboratories, it is mostly utilized for research investigations and epidemiological purposes. Patients can have anti-CHIKV antibodies found soon after the onset of symptoms; typically, IgM is found five days later, and IgG is found a few days later. Although there are commercial enzyme immunoassays and immunofluorescence tests, they perform poorly when used by experts [116]. Serological results should be interpreted with caution due to 1) the possibility of false negatives from CHIKV-induced mixed cryoglobulinemia [117], 2) cross-reactivity with viruses of the Semliki Forest serocomplex that need to be seroneutralized, and 3) the persistence of anti-CHIKV IgM for months after the onset of the disease. To show a recent CHIKV infection, it is usually enough to test a sample taken during the acute stage and another sample taken at least three weeks later. Any uncertainty should result in a request for help from a knowledgeable lab [118]. **(Table 1)**

| Diagnostic methods | Description | Efficacy | Advantages | Limitations | References |
|---|---|---|---|---|---|
| RT- PCR | Detection of viral RNA (vRNA) is | Highly effective | Highly specific, | Specialists and advanced | [119] |

| | important and specific for viruses. | during the acute phase. | rapid, and sensitive. | equipment are required. | |
|---|---|---|---|---|---|
| Antigen detection | Specifically, it detects viral proteins. | Less reliable than RT-PCR | It can be used in limited settings and rapid methods. | Lower specificity and sensitivity. | [120] |
| Serological testing | Detection of IgM during the first week of infection. | Cross-reaction may occur from moderate to high effectiveness. | Useful for retrospective diagnosis. | Cross-reaction may occur with other flaviviruses. | [121] |
| Virus isolation | The virus is cultured from patients' samples. | It is not routinely used due to the complexity of the process and the time required. | Highly specific and widely used for research purposes. | Specialized facilities and time is required. | [122] |

In vulnerable locations with current Aedes spp. Activity and public health depend on the effective management of patients with acute CHIKV infection. Within these regions, the majority of health authorities advise early detection of imported or indigenous illnesses, appropriate diagnostic tool use, isolation of suspect patients, quick communication with the local health department, and occasionally required case notification. The ultimate goal is to prevent epidemics from spreading after new occurrences [123].

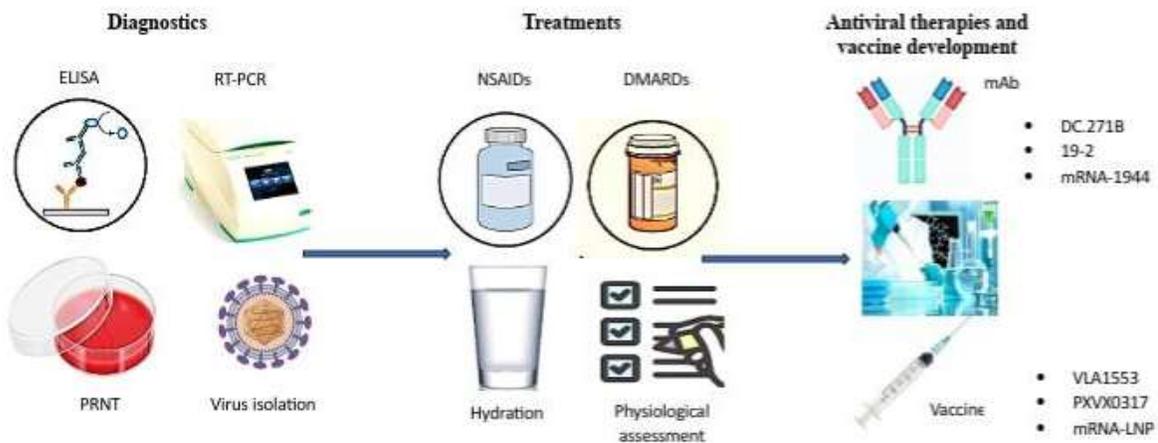

**Figure 5: Chikungunya disease management**

## 8. Complications:

Chronic arthritis is the serious and most common complication of CHIKF. Episcleritis, optic neuritis, retinitis, conjunctivitis, iridocyclitis, and uveitis are the main ocular manifestations of the disease [124]. Iridocyclitis and retinitis are the most common complications that are found to be benign and self-limited [125].

### 8.1. Consultation:

Chronic arthritis due to CHIKF has emerged as a concern for rheumatologist specialists globally due to recurring pain and swelling that poorly responds to analgesic treatment and because it is a debilitating disease that mimics rheumatic arthritis and comprises the quality of life [126].

### 8.2. Enhancing Healthcare team outcomes:

Mosquitoes are the arthropod vectors that impact the world the most; these are responsible for the transmission of diseases from CHIKV to malaria that are incalculable in terms of morbidity and mortality. Due to the lack of treatment for most of these diseases, vector control is the main means of prevention. In the past, insecticides were thought to be effective, but now, this widespread resistance has put a dent in these efforts [127]. (Level V) Another focus of prevention in Brazilian studies has been proven to be personal protection and individual actions, including the use of DEET, light-colored long-sleeve clothing, reduction of peri-domiciliary water poodles or reservoirs, and bed nets [128] (Level I).

When it comes to improving outcomes, the Centers for Disease Control (CDC) has highlighted the dangers of CHIKV and DENV confection and the differences in treatment modalities, making it important for all healthcare workers to avoid DENV-related deaths [129]. In 2018, an observation highlighted that there is a need for proper diagnosis (gathering serology DENV and Zv of the infected individuals), management, and surveillance to avoid poor prognosis [130] (Level I).

9. **Prevention:**

The only effective preventive strategies until a vaccine is developed are vector control and personal protection against mosquito bites. The same approach used to combat dengue has also been used to control adult and larval mosquito populations, with varying degrees of success in different nations and environments [131]. The most effective way to avoid CHIKV infection is to decrease mosquito populations. Breeding sites need to be eliminated, demolished, cleaned, and treated with pesticides on a regular basis [132]. Clothing that reduces skin exposure to day-biting vectors is recommended for protection. Apply repellents sparingly on exposed skin or clothing, following the directions on the product label. It is recommended that repellents include DEET (N, N-diethyl-3-methylbenzamide), Icaridin (1-piperidine carboxylic acid, 2-(2-hydroxyethyl)-1-methylpropylester), or IR3535 (3-[N-acetyl-N-butyl]-amino propionic acid ethyl ester). Another way to lessen indoor biting is by using pesticide vaporizers or mosquito coils [133].

10. **Vaccination:**

For CHIKV, there isn't a commercial vaccine available just yet, while some potential shots have been tried on humans [134]. A variety of technologies, including consensus-based DNA vaccines, recombinant subunit vaccines, alphavirus chimeras, live-attenuated viruses, inactivated viral vaccines, and, more recently, a vaccine based on virus-like particles (VLPs), have been employed to create CHIK vaccines. Phase I studies for two vaccine candidates are now complete: the VRC-

CHKVLP059-00-VP, VLP vaccine, and a live recombinant measles-virus-based chikungunya vaccine. The live recombinant measles virus-based chikungunya vaccine was safe, had a tolerability profile that was generally acceptable, and shown good immunogenicity even in the presence of measles immunity [135]. In addition, the VLP vaccine, VRC-CHKVLP059-00-VP, was well tolerated, safe, and immunogenic [136].

## 11. Treatment:

There is presently no licensed targeted therapy for acute CHIKV infection. Treatment is primarily supportive care and includes the use of analgesic and anti-inflammatory medication, rehydration, and rest. However, research to identify potential new antiviral therapies or repurposing of existing compounds for treating CHIKV infection is ongoing. For example, chloroquine has in vitro activity against several viruses, and has been found to inhibit CHIKV replication in Vero cells [137]. However, it has not been shown to have anti-CHIKV effects in vivo. Compounds that may interfere with viral entry, including phenothiazines [138] and flavaglines [139] are being investigated as potential therapies. In vitro CHIKV action has been demonstrated for ribavirin, which is used with doxycycline to lower viral loads and inflammation in infected mice. Following viral challenge, mice and nonhuman primates have been shielded against contracting CHIKV infection by [140] monoclonal antibodies to the E1 and E2 proteins [141]. It is uncertain, nevertheless, if hyper immune serum or monoclonal antibodies may be used to passively immunize against CHIKV illness once infection has already begun [142]. **(Table 2)**

| Treatment | Description | Effectiveness | Advantages | Limitations | References |
|---|---|---|---|---|---|
| NSAIDs | Anti-inflammatory and non-steroidal drugs that reduce join pain and inflammation. | Effectively help in symptoms management. | Reduce pain and are easily available. | Cause gastrointestinal problems in some people. | [143] |
| Paracetamol | Help in management of fever and mild pain. | Effectively help in symptoms management. | Well tolerated and generally safe. | Liver may damage due to overuse. | [144] |
| Oral rehydration solution (ORS) | Help in prevention of dehydration. | Essential in severe cases. | Easy to administer. | None significant if used correctly. | [145] |
| Physiotherapy | In chronic cases reduce discomfort and improve mobility. | Important for long term management. | Reduce chronic pain and enhance mobility. | Requires regular checkup and professional guidance. | [146] |
| Bed rest | Reduces fatigue and aids recovery. | Important for recovery. | The body heals and recovers energy. | Long-term rest may cause muscle stiffness. | [147] |

**12. Development of CHIKV antiviral therapies:**

Antiviral treatments and vaccines play a crucial role in addressing CHIKV infections, with numerous potential antiviral options currently being researched, as detailed by Battisti and others and Hucke and Bugert [148]. Recently, 3-methyltoxoflavin has been recognized as a blocker of CHIKV and other viruses, alongside itraconazole. These two effective inhibitors of CHIKV replication show great potential and may be repurposed as broad-spectrum antivirals. A deeper understanding of the molecular mechanisms behind viral replication and virulence is crucial for creating effective CHIKV-specific antiviral treatments [149]. Monoclonal antibodies (mAbs) aimed at CHIKV have been developed and tested in mice to assess safety, efficacy, dosing, and mechanisms of action prior to moving into human clinical trials. For example, the human mAb DC.271B, which targets the CHIKV E2 epitope, provided significant protection, whereas other mAbs directed at the CHIKV E1 epitope were less effective. Another monoclonal antibody, 19–1, which is an IgG2b κ-chain isotype targeting the CHIKV E2, exhibits strong binding affinity and high sensitivity to CHIKV [150]. A notable strategy from Moderna involves lipid nanoparticle-encapsulated messenger RNA-1944 (mRNA-1944), which encodes the heavy and light chains of a CHIKV-specific neutralizing antibody. A randomized phase I trial was conducted from January 2019 to June 2020 to assess its safety and pharmacological activity. Despite positive results from Phase I, Moderna chose not to advance this candidate to a Phase II clinical trial [151].

### 13. Vector surveillance and control:

Ultimately, addressing CHIKV and other arboviruses requires crucial efforts to monitor and manage mosquito vectors, which is essential for reducing disease transmission and protecting public health [152]. Tracking *Aedes* mosquito populations in endemic regions can aid in predicting and preparing for possible CHIKV outbreaks [153]. Additionally, population-based surveys and seroprevalence studies are valuable for evaluating the level of CHIKV exposure in communities and guiding public health strategies. Geographic Information Systems (GIS) technology is utilized to map CHIKV transmission patterns and pinpoint high-risk regions, helping to allocate resources for vector control and healthcare services [154]. Vector control strategies focus on decreasing the populations of disease-carrying mosquitoes to lower the risk of viral transmission [155].

In conclusion, addressing CHIKV and other *arboviruses* requires crucial efforts to monitor and manage mosquito vectors, which are essential for effectively reducing disease transmission and protecting public health. Keeping track of *Aedes* mosquito populations in endemic regions can assist in predicting and preparing for potential CHIKV outbreaks. Population-based surveys and seroprevalence studies are valuable for evaluating the level of CHIKV exposure in a community and guiding public health interventions [156, 157]. Geographic Information Systems (GIS) technology is employed to map CHIKV transmission patterns and identify high-risk areas, facilitating resource allocation for vector control and healthcare services. Vector control strategies focus on lowering the populations of disease-carrying mosquitoes to reduce the risk of viral transmission [158]. One effective strategy involves releasing Wolbachia-infected mosquitoes in different areas to decrease local *Aedes aegypti* populations, which helps lower the transmission of CHIKV and other *arboviruses*. Additionally, other vector control methods include the application of insecticides, larvicides, and environmental changes to eliminate mosquito breeding sites. Community involvement and education are essential for motivating individuals to adopt preventive measures, such as using mosquito nets and repellents and maintaining proper sanitation to lower the risk of infection. Continuous research and innovative strategies in vector control are vital in the battle against CHIKV and other *arboviruses*, aiming to lessen the impact of these diseases on public health [159].

### 14. Economic Burden:

The 2006 epidemic in India imposed a heavy epidemiological burden and productivity loss on the community. 25588 DALYs is the national recorded loss during the 2006 CHIKV epidemic. Arthralgia accounts for 69% of the total DALYs, showing a heavy burden. Productivity loss in terms of income was recorded at 6 million USD [160]. A study conducted in La Reunion in 2006 with military policemen reported that 93.7% of symptomatic patients complained of a chronic stage of the illness, which is characterized by pains in bones, joints, or both, although the study was conducted about six months after the epidemic peak. Most of the working agents were disabled due to the loss of mobility, hand disability, and depression reaction, which lasts for weeks to months and has negative consequences on social organization, health, and economy in epidemic regions [161, 162]. With these precedents, if the outbreak spreads throughout Mexico, the infected working individuals will be prostrated, which will increase the economic burden.

## 15. Challenges and Future Directions in CHIKV Research and Control:

Although there has been significant progress in the investigation of CHIKV pathogenesis, exploration of therapeutic and vaccine options, and development of diagnostic tools, several challenges, and future directions remain particularly relevant for the management of CHIKV infections and their reduction in impact. CHIKF shows symptoms like other arthropod vector diseases such as dengue fever and Zika virus [163]. Therefore, it is essential to develop new, rapid, and accurate differential diagnostic tests that can differentiate CHIKV from other arboviruses. Moreover, understanding the impact of coinfection with multiple arboviruses on disease severity and clinical manifestations is a growing point of concern [164]. The lack of specific antiviral therapies for CHIKV is a major problem in severe cases of the disease. Significant progress has been made in treatment during the last decade. However, ongoing research into antiviral drug development and repurposing of existing drugs may lead to more effective treatments [165].

Although CHIKV is a global health threat, mostly middle to low-income countries that have limitations to healthcare infrastructure or resources for outbreak response and surveillance are affected more by CHIKV [166]. Collaboration between governments, research institutions, international organizations, and public health agencies is necessary for sharing knowledge, coordinating outcomes, and to overcome the global impact of CHIKV, particularly in low-income countries. Additionally, efforts should be done to expand vaccines production and options, reduce costs, and establish distribution centers or other mechanisms that reach vulnerable communities, thus ensuring equal access to CHIKV vaccines. Effective vector control remains a cornerstone for CHIKV prevention [167]. *Aedes* mosquitoes are highly adaptable to temperate regions and their resistance to insecticides poses a global challenge [168]. So, development of alternative vector control methods such as Wolbachia-based strategies and genetically modified mosquitoes could prevent CHIKV future outbreaks [169]. Irreversible climate change is extending the geographic range of CHIKV that is enhancing CHIKV transmission to other new regions [170]. Research and global efforts should be taken to monitor and predict the impact of climate change on vector distribution.

## 16. Conclusion:

CHIKV is a continuous significant public health problem due to its potential for large scale outbreaks and the persistent challenges that it presents. Understanding of CHIKV virology, pathogenesis, transmission route, and the development of diagnostic strategies and vaccines are essential for mitigating its impacts. Due to the adaptability of *Aedes* mosquitoes to urban environments and the effects climate change on their distribution, vector control is the preferred prevention mean for controlling CHIKV transmission. Development of durable vaccines and antiviral therapies is the pressing need along with equal access to these interventions. Ongoing research must overcome challenges such as accurate diagnosis that differentiate between arbovirus infections, co-infection, and the long term efficacy of vaccines. International collaboration, proactive surveillance, and robust public health agencies are vital for early detection and effective control of CHIKV epidemics.